\documentclass[10pt,onecolumn,draftclsnofoot]{IEEEtran}
\usepackage{algorithm}
\usepackage{algorithmicx}

\usepackage{amssymb,amsmath} %for math symbols and etc 
\usepackage{graphicx}        %for graphics 
\usepackage{float}  		%for float graphics, i.e. , I can place graphics how I want
\usepackage{url}
\usepackage{float} 
\usepackage{psfrag} 
\usepackage{url}
\usepackage{subfig}

\newtheorem{problem}{Problem}
\usepackage{chemarr}
\usepackage{fancybox}
\usepackage{multicol}
\usepackage{wrapfig}
\usepackage{stfloats}
\usepackage{mathtools, cuted}
\usepackage{url}
\usepackage{cite}
\usepackage{empheq}
\usepackage{epstopdf}
\usepackage{xcolor}
\usepackage{tabularx}
\usepackage{url}
\usepackage{hyperref}

\usepackage[margin={1.60cm,1.60cm}]{geometry}

\usepackage{tikz}
\newcommand{\boxalign}[2][0.97\textwidth]{
  \par\noindent\tikzstyle{mybox} = [draw=black,inner sep=6pt]
  \begin{center}\begin{tikzpicture}
   \node [mybox] (box){%
    \begin{minipage}{#1}{\vspace{-2mm}#2}\end{minipage}
   };
  \end{tikzpicture}\end{center}
}

\begin{document}

%\title{Observing the states of nonlinear networks}

\title{Joint Sensor Node Selection and State Estimation for Nonlinear Networks and Systems}

\author{Aleksandar Habe$\text{r}^1$
	\thanks{
		$^1$Department of Engineering and Environmental Science, The City University of New York, College of Staten Island, 2800 Victory Blvd, New York, NY 10314, USA. 	Email: aleksandar.haber@csi.cuny.edu.}
		% This work is supported by the PSC-CUNY Award A (61303-00 49,62267-00 50), the PSC-CUNY Award B (63761-00 51).}
}

%\markboth{Journal of \LaTeX\ Class Files,~Vol.~xx, No.~x, January~xxxx}%

\maketitle

\begin{abstract}
State estimation and sensor selection problems for nonlinear networks and systems are ubiquitous problems that are important for the control, monitoring, analysis, and prediction of a large number of engineered and physical systems. Sensor selection problems are extensively studied for linear networks. However, less attention has been dedicated to networks with nonlinear dynamics. Furthermore, widely used sensor selection methods relying on structural (graph-based) observability approaches might produce far from optimal results when applied to nonlinear network dynamics. In addition, state estimation and sensor selection problems are often treated separately, and this might decrease the overall estimation performance. To address these challenges, we develop a novel methodology for selecting sensor nodes for networks with nonlinear dynamics. Our main idea is to incorporate the sensor selection problem into an initial state estimation problem. The resulting mixed-integer nonlinear optimization problem is approximately solved using three methods. The good numerical performance of our approach is demonstrated by testing the algorithms on prototypical Duffing oscillator, associative memory, and chemical reaction networks. The developed codes are available online. 
\end{abstract}


\IEEEpeerreviewmaketitle

%%%%%%%%%%%%%%%%%%%%%%%%%%%%%%%%%%%%%%%%%%%%%%%%%%%%%%%%%%%%%%%%%%%%%%%%%%%%%%%%%%%%%%%%%%%%%%%%%%%%%%%%%%%%%%%%

\section{Introduction}
In a large variety of engineering and scientific fields, we are often faced with the problem of estimating states of networks with nonlinear dynamics. For example, this problem is crucial for identification, estimation, monitoring, and control of power systems, communication, traffic, biochemical, biophysical, combustion reaction, and ecological networks, as well as for other systems with nonlinear dynamics~\cite{motter2015networkcontrology,pang2017universal,liu2011controllability,liu2013observability,angulo2019theoretical,haber2014subspace,haber2018sparsity,haber2016sparse,haber2017state,haber2013moving,haber2012identification}. 

Generally speaking, state estimation for nonlinear networks consists of two steps. In the first step, which is often referred to as the \textit{sensor node selection step}, we are interested in selecting a subset of network nodes whose states or output variables should be observed such that from this limited information we can accurately reconstruct the global network state. In the second step, which is often referred to as the \textit{observer design step} or \textit{state reconstruction step}, we are interested in designing an algorithm (observer) to reconstruct the global network state using the information collected from the sensor nodes assigned in the first step.  The estimated states might be used for system monitoring, prediction, or control. Control node selection and control action design are dual problems to the estimation problems, for more details see~\cite{pasqualetti2014controllability,summers2015submodularity,letellier2018nonlinear,aguirre2018structural,haber2017state}.

There is a large body of literature on sensor/control node selection for network dynamics. A detailed literature survey of all methods and approaches goes well beyond the scope and length limits of this manuscript. Consequently, in the sequel, we mention the main lines of research and recent contributions. Widely used and arguably most popular approaches for sensor (control) node selection are relying on graph-based methods stemming from structured control theory~\cite{reinschke1988multivariable,lin1974structural,siljak2011decentralized}. These approaches have been used in~\cite{liu2011controllability,liu2013observability} and in a large number of follow-up contributions, to devise methods for control and sensor node selection for complex networks. For example, the main idea of the approach proposed in~\cite{liu2013observability}, is to select sensor nodes by searching for strongly connected components in a graph describing node connections. This is a computationally inexpensive approach that can produce initial sensor node locations that can be used to initialize optimization-based sensor selection algorithms~\cite{haber2017state}. However, several authors have investigated and highlighted some of the drawbacks of graph-based approaches~\cite{pasqualetti2014controllability,summers2015submodularity,letellier2018nonlinear,aguirre2018structural}. Namely, in some cases, the graph-based approaches might produce far from optimal solutions, or selected sensor nodes might lead to numerically ill-conditioned estimation problems~\cite{haber2017state}. Recent structural controllability and observability approaches, as well as recent surveys on these topics, can be found in~\cite{doostmohammadian2020recovering,angulo2019structural,kahl2019structural,olshevsky2020relaxation,klickstein2018control}.

Another line of research is to select sensor (control) nodes for linear networks by optimizing suitable observability (controllability) performance criteria depending on system Gramians, see for example~\cite{summers2015submodularity,pasqualetti2014controllability} and follow-up approaches. Such methods are mainly designed for linear network dynamics and might not be applicable to the nonlinear case. Empirical Gramians~\cite{lall2002subspace} of nonlinear systems have been used in ~\cite{qi2014optimal,serpas2013sensor} to select actuators (control nodes) and sensor nodes for nonlinear networks and systems. The main limitation of these approaches is that the computation of empirical Gramians is a computationally challenging task even for small-sized networks. In~\cite{haber2017state}, we have developed a sensor selection approach for nonlinear networks with smooth dynamics. However, the main prerequisite for applying this method and empirical Gramian approaches is \textit{a priori} information on the initial state for which observability/controllability metrics are defined. In practice, we can rarely obtain accurate \textit{a priori} information on the network initial state. An approach for sensor selection of nonlinear networks has been presented in~\cite{nugroho2019sensor}. This method assumes a specific form of nonlinear dynamics and its applicability to a broader class of nonlinear networks has to be further investigated. In~\cite{letellier2018symbolic,sendina2019observability,letellier2018nonlinear}, approaches have been developed to tackle some of the drawbacks of pure graph-based observability methods. These methods rely on a symbolic network Jacobian matrix and on symbolic observability coefficients. In principle, these methods can be coupled with nonlinear estimators to devise a complete solution to the network estimation problem. However, this idea needs to be numerically validated. Sensor and actuator placement problems for linear systems as well as for systems described by partial differential equations have been considered in~\cite{lou2003optimal,morris2010linear,edalatzadeh2019optimal2,edalatzadeh2019optimal,zare2019proximal}. 

To the best of our knowledge, sensor selection and state estimation problems are usually treated separately. The widely used approach is to first select sensor nodes by optimizing an observability criterion and then to design an estimator. Our numerical experience and results show that for certain classes of nonlinear networks this approach might lead to far from optimal estimation performance. \textit{Our hypothesis is that by treating the sensor node selection and state estimation problems as a joint problem, we can improve the overall estimation performance.}

In this manuscript, we propose a novel problem formulation and novel methods for joint state estimation and sensor selection for nonlinear network dynamics. Our main idea is to incorporate the sensor selection problem into an initial state estimation problem. The resulting Mixed-Integer Nonlinear Optimization (MINO) problem is approximately solved using three methods. The first method uses the Mesh Adaptive Direct Search (MADS) algorithm~\cite{le2011algorithm} to solve the MINO problem. This is a derivative-free optimization method requiring only a procedure to compute the cost function. In an initial phase, the second and third methods solve a relaxed MINO problem obtained by replacing the integer constraints with constraints involving continuous variables. Then, the second and third methods use the solution of the relaxed problem to formulate and solve Mixed-Integer Linear Programming (MILP) problems. These MILP problems are solved by using a hybrid branch and bound method. The second and third methods are partly inspired by approaches for solving mixed-integer optimization problems arising in optimal predictive control~\cite{sager2009reformulations,sager2011combinatorial,burger2019design}.

The key contribution of this article is a novel unified framework for joint sensor node selection and initial state estimation. Furthermore, we devise and thoroughly numerically test three approaches for approximating the solutions of sensor selection problems for nonlinear systems. The methods developed in this paper approximately solve an objectively difficult optimization problem. Namely, in its full generality, the considered MINO problem is constrained, NP-hard, and non-convex. An additional layer of complexity to this problem comes from the fact that the optimization problem is constrained by nonlinear dynamics. Furthermore, all three developed methods aim at computing an approximate solution of the MINO problem by relaxing the integer constraints or by direct search. Due to all these layers of complexity, it is challenging to derive a generally-applicable convergence analysis of the developed methods. Thorough theoretical analysis might be possible only for relatively simple forms of nonlinear dynamics. To provide valuable insights into the behavior of the developed approaches, we investigate the performance of the methods by performing extensive simulation experiments. The good numerical performance of our methods is demonstrated by testing them on Duffing oscillator networks, associative memory networks, and on two models of chemical reaction networks. We compare our methods with exhaustive search and random selection of sensor nodes. Compared to other methods for solving MINO problems~\cite{belotti2013mixed}, such as methods based on branch and bound algorithms, our approach is relatively easy to implement, does not rely on convexification procedures that are usually case dependent, and is applicable to a broad class of smooth nonlinear systems. The developed codes are available online~\cite{haberSensorCode2020}. 

This manuscript is organized as follows. In Section~\ref{problemFormulation}, we present the problem formulation. In Section~\ref{sensorSelectionAlgorithms}, we present the sensor selection algorithms. In Section~\ref{gradientsAndGeneralizations} we derive expressions for the derivatives of cost functions used in this paper and present some generalizations of the developed approaches. Numerical results and conclusions are presented in Section~\ref{numericalResults} and Section~\ref{conclusions}, respectively.

\section{Problem Formulation}
\label{problemFormulation}

In this section, we incorporate the sensor selection problem into an open-loop initial state estimation problem. This enables us to formulate the sensor selection and initial state estimation problems as a single MINO problem. Three algorithms for approximating the solution of this problem are stated in Section~\ref{sensorSelectionAlgorithms}. 

First, we explain the used notation. Let $\mathbf{w}_{k}\in \mathbb{R}^{n}$ denote an arbitrary $n$-dimensional discrete-time vector at a discrete-time instant $k\in \mathbb{Z}_{0}^{+}$. The notation $\mathbf{w}_{0:S}$ denotes a lifted column vector $\mathbf{w}_{0:S}: =\text{col}(\mathbf{w}_{0},\mathbf{w}_{1},\ldots, \mathbf{w}_{S}),\; \mathbf{w}_{0:S}\in \mathbb{R}^{(S+1)n}$, where $S\in \mathbb{Z}_{0}^{+}$. The notation  $\text{col}(\mathbf{w}_{0},\mathbf{w}_{1},\ldots, \mathbf{w}_{S})$ denotes an operator defined by $\text{col}(\cdot)=[\mathbf{w}_{0}^{T},\mathbf{w}_{1}^{T},\ldots, \mathbf{w}_{S}^{T}]^{T}$. The notation  $\mathbf{g}[\mathbf{w}]$ denotes a vector function of the vector argument $\mathbf{w}$. The notation $\text{diag}(\mathbf{w})\in \mathbb{R}^{n\times n}$ denotes a diagonal matrix with the entries of the vector $\mathbf{w}$ on its main diagonal. The notation $I_{S}$ denotes an $S\times S$ identity matrix. 

We consider a nonlinear state-space model
\begin{align}
& \dot{\mathbf{x}}=\mathbf{f}[\mathbf{x}], \label{nonlinearStateSpaceModel} \\
& \mathbf{y}=C\mathbf{x},  \label{nonlinearOutputEquation}
\end{align}
where $\mathbf{x}\in \mathbb{R}^{Nn}$ is a global network state, $\mathbf{f}[\cdot]:R^{Nn}\rightarrow R^{Nn} $ is a nonlinear vector function that is sufficiently smooth, $\mathbf{y}\in \mathbb{R}^{M}$ is an observed global output vector, and $C\in \mathbb{R}^{M\times Nn}$ is an output matrix. We assume that the network consists of $N$ nodes, where a local state of each node is $n$-dimensional, that is, $\mathbf{x}=\text{col}(\mathbf{x}^{(1)},\mathbf{x}^{(2)},\ldots, \mathbf{x}^{(N)})$, where $\mathbf{x}^{(i)}\in \mathbb{R}^{n}$, $i=1,\ldots, N$, is the local state of the $i$-th node. We assume that the output vector $\mathbf{y}$ consists of $M$ entries $y^{(j)}\in \mathbb{R}$, where the index $j$ can take any value from the set $\{1,2,\ldots N \}$. The entry $y^{(j)}$ is a local observation at a prescribed node. A node is a \textit{sensor node} if its output is being observed. For presentation clarity, we have assumed a specific form of the output equation \eqref{nonlinearOutputEquation}. The approaches presented in this paper can be generalized to other nonlinear forms of the output equation. The integer $M$ denotes the number of sensor nodes. Next, let $\mathbf{z}=\text{col}(y^{(1)}, y^{(2)},\ldots, y^{(N)} )$, $\mathbf{z}\in \mathbb{R}^{N}$, denote an output vector obtained assuming that all network nodes are sensor nodes. We assume that the sensor node outputs are available at discrete time samples $t=kh$, where $k=0,1,2,\ldots,$ and $h$ is a discretization constant. Furthermore, let $\mathbf{y}_{k}:=\mathbf{y}[kh]$, $\mathbf{z}_{k}:=\mathbf{z}[kh]$, and $\mathbf{x}_{k}:=\mathbf{x}[kh]$. 

The locations of the sensor nodes are encoded in the structure of the matrix $C$. Knowing the model of the system dynamics, given by the equation \eqref{nonlinearStateSpaceModel}, our goal is to determine an optimal (sparsity) structure of the matrix $C$, under the condition that sensors can be placed on a fixed number of nodes that is usually smaller than $N$. At the same time, we want to ensure that the selected matrix $C$ will produce accurate results when used to design an estimator or an observer. We consider the sensor selection and initial state estimation problems under the following scenarios. In the first scenario, we can simulate the model of the network dynamics \eqref{nonlinearStateSpaceModel} for a user-selected initial condition. In this way, we can obtain the state trajectory $\{\mathbf{x}_{k}\}$, and consequently, we can construct the sequence $\{\mathbf{z}_{k}\}$. Then, using the sequence $\{\mathbf{z}_{k}\}$, and the model \eqref{nonlinearStateSpaceModel}, under the condition that only $M_{\text{max}}$ of the sensor nodes can be used, the goal is to select the optimal structure of the matrix $C$ such that the initial state of the real system can be accurately reconstructed. After we have determined the sensor locations, we can place sensors on predetermined locations, and these sensors can be used to collect the output sequence $\{\mathbf{y}_{k} \}$ from the real system that can be used for state estimation. In the second scenario, instead of simulating the system response (for a selected initial condition), we can place sensors on all nodes of the real system to obtain the sequence $\{\mathbf{z}_{k}\}$. Then, using this sequence collected from the real system, we can find the optimal locations of $M_{\text{max}}$ sensors. After completing the sensor selection procedure, all the sensors except $M_{\text{max}}$ selected sensors should be removed from the network, and the initial state should be estimated. This scenario is only feasible for small or medium-sized networks due to practical and economical constraints. It should be emphasized that in the first scenario, the user-selected initial condition necessary to obtain the data sequence $\{\mathbf{z}_{k}\}$ is usually different from the unknown initial conditions of the real physical system. There are two approaches to minimize this uncertainty. The first approach is to choose the initial conditions to be in the expected range of the initial conditions of the system. The second approach is discussed in Section~\ref{generalizationSubsections}. 

We can formally define the joint problem of selecting the sensor nodes and estimating the initial states as follows.

\begin{problem}[Joint Sensor Selection and Initial State Estimation]
Given the maximal number of sensor nodes, denoted by $M_{\text{max}}$, $M_{\text{max}}\le N$, and given the length of the observation horizon, denoted by $L$, determine:
\begin{enumerate}
\item Locations of sensor nodes using the sequence of data samples $\{ \mathbf{z}_{0}, \mathbf{z}_{1}, \ldots, \mathbf{z}_{L} \}$. These data samples can be obtained by using the two approaches. In the first approach, the data samples are obtained by simulating the system \eqref{nonlinearStateSpaceModel} for the user-selected initial condition. In the second approach, the samples are collected by placing sensors on all nodes of the real system. 
\item  An estimate of the real system initial state  $\mathbf{x}_{0}$ determined from the sequence of output data samples $\{ \mathbf{y}_{0}, \mathbf{y}_{1}, \ldots, \mathbf{y}_{L}  \}$ collected from the real system by the assigned sensor nodes.
\end{enumerate}
\end{problem}
To mathematically formulate the sensor selection problem, we represent the continuous-time dynamics~\eqref{nonlinearStateSpaceModel} in the discrete-time domain~\cite{alessandri2008moving,alessandri2003receding,moraal1995observer,haber2017state}. The simplest discretization approach is based on the Forward Euler (FE) method~\cite{iserles2009first}:
\begin{align}
\mathbf{x}_{k}=\mathbf{x}_{k-1}+ h\mathbf{f}[\mathbf{x}_{k-1}],
\label{FEdiscretizedDynamics}
\end{align}
The FE dynamics \eqref{FEdiscretizedDynamics} approximates relatively well the continuous-time dynamics when $h$ is small and when the dynamics is not stiff. Stiff network dynamics are characterized by time constants of local nodes that significantly differ in magnitude. A large number of systems such as reaction-diffusion systems, chemical reaction networks, and other systems coupling multi-physics phenomena have stiff dynamics. To allow for larger values of the discretization constant and to be able to accurately represent stiff network dynamics, we employ the Trapezoidal Implicit (TI) discretization method~\cite{iserles2009first}:
\begin{align}
\mathbf{x}_{k}=\mathbf{x}_{k-1}+0.5h \big(\mathbf{f}[\mathbf{x}_{k}]+ \mathbf{f}[\mathbf{x}_{k-1}] \big).\label{TEdiscretizedDynamics}
\end{align}
Another option is to use the implicit Runge-Kutta method~\cite{iserles2009first,haber2017state}, however, for brevity, we do not use such a method in this manuscript. To solve the sensor selection method we have to simulate the system dynamics many times. The main disadvantage of the TI method over the FE method is that in every simulation step $k$, we need to solve the system of nonlinear equations \eqref{TEdiscretizedDynamics} for $\mathbf{x}_{k}$. This generally results in $\mathcal{O}\big( N^{3}n^{3} \big)$ computational complexity. In sharp contrast, in every step $k$, the computational complexity of the FE method is $O(Nn)$. This implies that the methods presented in the next section will have a larger computational complexity for the TI dynamics. For notational brevity, both FE and TI discretized dynamics are denoted by the following equation 
\begin{align}
\mathbf{x}_{k}=\mathbf{g}\big[\mathbf{x}_{k},\mathbf{x}_{k-1}\big]. 
\label{uniqueRepresentation}
\end{align}
Next, we introduce a parametrized output equation relating states and  $\mathbf{z}_{k}$:
\begin{align}
\mathbf{z}_{k}= C_{\boldsymbol{\theta}}[\boldsymbol{\theta}]\mathbf{x}_{k},
\label{parametrizedOutputEquation}
\end{align}
where $\boldsymbol{\theta}\in \{0,1 \}^{N}$ is a binary parametrization vector and  $C_{\boldsymbol{\theta}}[\boldsymbol{\theta}]\in \mathbb{R}^{N\times Nn}$ is a parametrized output matrix. Depending on the structure of a system used for performing numerical experiments,  we assume two parmetrization forms of $C_{\boldsymbol{\theta}}[\boldsymbol{\theta}]$. For networks for which $n=1$, we assume $C_{\boldsymbol{\theta}}=\text{diag}\big(\boldsymbol{\theta} \big) $. For $n>1$, $C_{\boldsymbol{\theta}}$ is a block diagonal matrix with the $i$-th block equal to $C_{i}=[\theta_{i}\; 0 \; \ldots \; 0]\in \mathbb{R}^{1\times n}$, where $\theta_{i}$ is the $i$-th entry of $\boldsymbol{\theta}$. Our idea for developing the sensor selection method originates from the observability definition for discrete-time systems~\cite{hanba2009uniform,haber2017state}. Namely, the observed output sequence $\mathbf{y}_{0:L}$ and the initial state $\mathbf{x}_{0}$ are related by a nonlinear function $\mathbf{w}$
\begin{align}
\mathbf{y}_{0:L}=\mathbf{w}\big[ \mathbf{x}_{0}  \big],
\label{observabilityDiscreteTime1}
\end{align}
where the notation $\mathbf{y}_{0:L}$ is used to denote a lifted vector (see the beginning of Section~\ref{problemFormulation} for used notation) and the nonlinear function $\mathbf{w}[ \mathbf{x}_{0} ] : \mathbb{R}^{Nn}\rightarrow \mathbb{R}^{(L+1)M}$ is defined by 
\begin{align}
\mathbf{w}[ \mathbf{x}_{0} ] := \big(I_{L+1}\otimes C \big) \mathbf{x}_{0:L} =  \big(I_{L+1}\otimes C \big) \mathbf{x}_{0:L}[\mathbf{x}_{0} ],
\label{wFunction}
\end{align}
where $I_{L+1}\in\mathbb{R}^{(L+1)\times (L+1)}$ is an identity matrix and $\otimes$ denotes the Kronecker product. The state sequence $ \mathbf{x}_{0:L}[\mathbf{x}_{0} ]$ is only a function of the initial state $\mathbf{x}_{0}$ since from \eqref{uniqueRepresentation} we obtain telescopic equalities
\begin{align}
\mathbf{x}_{1}  & =\mathbf{g}[\mathbf{x}_{1},\mathbf{x}_{0}],\;\; \mathbf{x}_{2}=\mathbf{g}[\mathbf{x}_{2},\mathbf{x}_{1}], \;  \ldots \;, \label{telescopicEqualities1}  \\
\mathbf{x}_{L}  & =\mathbf{g}[\mathbf{x}_{L},\mathbf{x}_{L-1}]. \label{telescopicEqualitiesL}  
\end{align}
Consequently, for known $\mathbf{x}_{0}$ we can compute $\mathbf{x}_{k}$, $k=1,2,\ldots,L$, by solving the nonlinear equations (for the TI dynamics) or by simply propagating the telescopic equations (for the FE dynamics). Let us recall the \textit{uniform observability definition}. A discrete-time dynamical system is uniformly observable on a set if there exists $L>0$ such that $\mathbf{w}[ \mathbf{x}_{0}]$ is an injective function with respect to initial state $ \mathbf{x}_{0}$~\cite{hanba2009uniform}. From the practical point of view, this means that the system is observable if we can uniquely solve the system of nonlinear equations~\eqref{observabilityDiscreteTime1} for $\mathbf{x}_{0}$. This problem can be reformulated as a nonlinear optimization problem
\begin{align}
&\min_{\mathbf{x}_{0}}  \left\|  \mathbf{y}_{0:L}-\mathbf{w}\big[ \mathbf{x}_{0}  \big]  \right\|_{2}^{2}, \;\; \text{subject to \eqref{telescopicEqualities1}-\eqref{telescopicEqualitiesL}.}
\label{nonlinearOptimizationProblem}
\end{align}

 Similarly to the continuous-time nonlinear systems, sufficient observability conditions for discrete-time dynamics can be formulated. The uniform observability condition is that the Jacobian matrix of $\mathbf{w}[ \mathbf{x}_{0}]$ is of full-rank~\cite{hanba2009uniform,haber2017state}. In~\cite{haber2017state} we have developed a sensor selection method that optimizes a criterion that involves the Jacobian matrix. However, the Jacobian matrix is defined locally for a specific initial state $\mathbf{x}_{0}$, and consequently, the observability condition will vary from one state to another. Also, in practice the initial state knowledge is limited, and consequently, the sensor selection approach that is solely based on optimizing a criterion depending on the Jacobian matrix might produce inaccurate results. \textit{This motivates us to incorporate the sensor selection problem into the state estimation problem~\eqref{nonlinearOptimizationProblem} and to jointly solve the resulting problem for the initial state and sensor locations.} Following this idea, we formulate the joint problem of selecting the sensor nodes and estimating the initial state as the solution of the following MINO problem:
\begin{subequations}\label{problemPInteger}
{
%\begin{empheq}[box=\fbox]{align}
\boxalign[0.43\textwidth]{
\begin{align}
&(\textbf{P1})\;\min_{\mathbf{x}_{0},\boldsymbol{\theta}}  \;\;\;\left\|\mathbf{z}_{0:L}- \big(I_{L+1}\otimes C_{\boldsymbol{\theta}}[\boldsymbol{\theta}] \big) \mathbf{x}_{0:L}[\mathbf{x}_{0}]  \right\|_{2}^{2},
\label{OptimizationProblem1} \\
&\text{subject to} \;\;  \mathbf{x}_{i}=\mathbf{g}[\mathbf{x}_{i},\mathbf{x}_{i-1}], \; i=1,2,\ldots, L,  \label{constraint1} \\
& \sum_{l=1}^{N} \theta_{l} \le M_{\text{max}} ,  \Big(  \sum_{l=1}^{N} \theta_{l} = M_{\text{max}}   \Big), \label{constraint201}\\
&  \;\theta_{l}\in \{0,1 \},\; l=1,2,\ldots,N, \label{constraint2}   \\
& \underline{\mathbf{x}}_{0} \le \mathbf{x}_{0} \le \bar{\mathbf{x}}_{0}. \label{constraint3}
\end{align}
}
%\end{empheq}
}
\end{subequations}
The constraint \eqref{constraint201} limits the number of selected sensor nodes. We have two options for incorporating these constraints into the optimization problem. The first option, represented by the $\le$ relation, ensures that the number of selected nodes is \textit{smaller than or equal} to $M_{\text{max}}$. The second option, represented by the equality relation (in the brackets), is used to ensure that the number of sensor nodes is precisely \text{equal} to $M_{\text{max}}$. The second option is introduced since in our simulations we want to compare the developed algorithms with random sensor selection, and in order to ensure that the comparison is fair, we need to ensure that the selected number of sensor nodes is always constant. Also, we noticed that nonlinear solvers occasionally perform better in the case of the second option. In \eqref{constraint3} the notation $ \le$ denotes the element-wise less than equal relation, $\underline{\mathbf{x}}_{0}\in \mathbb{R}^{Nn}$ and $\bar{\mathbf{x}}_{0}\in \mathbb{R}^{Nn}$ are lower and upper bounds on $\mathbf{x}_{0}$. By relaxing the binary constraints and by eliminating the constraints on the maximal number of sensor nodes, from the MINO problem \eqref{problemPInteger} we obtain a relaxed problem:

\begin{subequations}\label{problemPIrelaxed}
{
%\begin{empheq}[box=\fbox]{align}
\boxalign[0.43\textwidth]{
\begin{align}
&(\textbf{P2})\;\min_{\mathbf{x}_{0},\boldsymbol{\theta}}  \;\;\;\left\|\mathbf{z}_{0:L}- \big(I_{L+1}\otimes C_{\boldsymbol{\theta}}(\boldsymbol{\theta}) \big) \mathbf{x}_{0:L}[\mathbf{x}_{0}]  \right\|_{2}^{2},
\label{OptimizationProblem1Relaxed} \\
&\text{subject to}\;\; \mathbf{x}_{i}=\mathbf{g}[\mathbf{x}_{i},\mathbf{x}_{i-1}], \; i=1,2,\ldots, L,  \label{constraint1Relaxed} \\
& \mathbf{0} \le \boldsymbol{\theta} \le \mathbf{1}, \label{constraint2Relaxed}  \\
& \underline{\mathbf{x}}_{0} \le \mathbf{x}_{0} \le \bar{\mathbf{x}}_{0}. \label{constraint3Relaxed}
\end{align}
}
%\end{empheq}
}
\end{subequations}
In \eqref{problemPIrelaxed},  $\mathbf{0}\in \mathbb{R}^{N}$ and $\mathbf{1}\in \mathbb{R}^{N}$ are the vectors of zeros and ones, respectively. 

\section{Sensor Selection Methods}
\label{sensorSelectionAlgorithms}
In this section, we introduce three algorithms for approximating the solution of the MINO problem \textbf{P1} in \eqref{problemPInteger}. The first algorithm is based on solving the problem using the NOMAD algorithm~\cite{currie2012opti}. The NOMAD algorithm is initialized with a solution guess generated by solving the relaxed problem \textbf{P2} defined in \eqref{problemPIrelaxed}. The second and third algorithms are based on computing the initial solution guess by solving the relaxed problem and then using this initial solution to formulate and solve MILPs that are defined in the sequel. The developed codes are available online~\cite{haberSensorCode2020}.

\subsection{Algorithm 1 - Mesh Adaptive Direct Search Approach}

This method is summarized in Algorithm~\ref{algorithm1}. The NOMAD algorithm (also known as  MADS) is a derivative-free optimization method. We use a version of this algorithm implemented in the OPTI MATLAB toolbox~\cite{currie2012opti}. The advantage of this approach is that it can easily be integrated with all MATLAB nonlinear solvers that are necessary to solve the system of equations~\eqref{constraint1}. Furthermore, to implement this method we only need a numerical procedure to evaluate the cost function.

In step 1 of Algorithm~\ref{algorithm1}, we solve the relaxed problem \textbf{P2} defined in  \eqref{problemPIrelaxed}. This problem is solved using the interior point method implemented in the MATLAB function $\text{fmincon}(\cdot)$. We use a recursive approach for solving the relaxed problem. The recursive approach is originally used for solving model predictive control problems, see Chapter 10 in \cite{lars2011nonlinear}. However, with minor modifications, it is also applicable to our problem. The recursive approach does not consider intermediate states $\mathbf{x}_{1},\mathbf{x}_{2},\ldots, \mathbf{x}_{L}$ in \eqref{constraint1Relaxed} as explicit optimization variables. Instead, in every iteration, the state sequences are computed by either forward propagation of the discrete-time dynamics \eqref{FEdiscretizedDynamics} (in the case of the FE method) or by solving the nonlinear system of equations \eqref{telescopicEqualities1}-\eqref{telescopicEqualitiesL} (in the case of the TI method). The nonlinear system of equations is solved using the trust-region dogleg method implemented in the MATLAB function $\text{fsolve}(\cdot)$. The solution of the relaxed problem is used in step 2 of Algorithm~\ref{algorithm1} as an initial guess for the NOMAD solver. By generating the initial guess in this way, we significantly decrease the number of iterations of the NOMAD algorithm. To implement the NOMAD algorithm, we use the previously explained recursive approach. Once the state sequence is computed, we can evaluate the cost function \eqref{OptimizationProblem1}, and this value is given as an input to the NOMAD solver. Finally, in step 3, using the sensor nodes computed by the NOMAD algorithm, we form the matrix $\hat{C}$ by eliminating the zero rows (corresponding to sensor nodes that are not selected) of the matrix $C_{\boldsymbol{\theta}}[\hat{\boldsymbol{\theta}}] $. For this $\hat{C}$, we form and solve the optimization problem \eqref{nonlinearOptimizationProblem} to compute the initial state estimate $\hat{\mathbf{x}}_{0}$. This problem is solved using the quasi-Newton method implemented in the MATLAB function $\text{fminunc}(\cdot)$.

	\begin{algorithm}[h]
		\caption{Sensor Selection Using NOMAD Method}\label{algorithm1}
%		\DontPrintSemicolon
				\textbf{inputs:} In the first phase (steps 1-2), the inputs are the output sequence $\{\mathbf{z}_{0}, \mathbf{z}_{1}, \ldots, \mathbf{z}_{L}  \}$ and $M_{\text{max}}$. In the second phase (step 3), the input is the output sequence $\{\mathbf{y}_{0}, \mathbf{y}_{1}, \ldots, \mathbf{y}_{L}  \}$ collected by the assigned sensor nodes. \\
		\textbf{outputs:} The optimal sensor selection vector $\hat{\boldsymbol{\theta}}$ and the initial state estimate $\hat{\mathbf{x}}_{0}$. \\
		\textbf{1. initial solution:} Solve the relaxed problem \textbf{P2} in \eqref{problemPIrelaxed}. Let the solution of this problem be denoted by $(\tilde{\boldsymbol{\theta}},\tilde{\mathbf{x}}_{0})$. \\
		\textbf{2. solve:} Using $(\tilde{\boldsymbol{\theta}},\tilde{\mathbf{x}}_{0})$ as an initial guess, solve the MINO problem \textbf{P1} in \eqref{problemPInteger} by using the NOMAD solver. Let the solution of this problem be denoted by  $(\hat{\boldsymbol{\theta}},\tilde{\mathbf{x}}_{0}^{(1)})$. \\
		\textbf{3. solve:} Implement the sensor locations on the physical system. Using the installed sensors, collect the output sequence $\{\mathbf{y}_{0}, \mathbf{y}_{1}, \ldots, \mathbf{y}_{L} \}$. Form the matrix $\hat{C}$ by eliminating the zero rows of the matrix $C_{\boldsymbol{\theta}}(\hat{\boldsymbol{\theta}})$ in \eqref{parametrizedOutputEquation}. Setting $C := \hat{C}$, solve \eqref{nonlinearOptimizationProblem} to compute $\hat{\mathbf{x}}_{0}$.	
\end{algorithm}
In Section~\ref{gradientsAndGeneralizations} we derive the gradients of the defined cost functions and we present some generalizations of the developed approaches.

\subsection{Algorithm 2 - Solving Relaxed and MILP Problems}

Here we present an algorithm for approximating the solution of the MINO problem \textbf{P1} in \eqref{problemPInteger} that is based on solving the relaxed problem \textbf{P2} in \eqref{problemPIrelaxed} and a MILP that is defined in the sequel. Let the solution of the relaxed problem be denoted by $\tilde{\mathbf{x}}_{0}$. Using this solution, we define the following cost function

\begin{align}
& J[\boldsymbol{\theta}]=\left\|  G[\boldsymbol{\theta}]  \right\|_{1}, \label{costFunction}\\
& G[\boldsymbol{\theta}]= \mathbf{z}_{0:L}- \big(I_{L+1}\otimes C_{\boldsymbol{\theta}}[\boldsymbol{\theta}] \big) \mathbf{x}_{0:L}[\tilde{\mathbf{x}}_{0}] \label{costFunction222}. 
\end{align}
The state sequence $\{\tilde{\mathbf{x}}_{1},\tilde{\mathbf{x}}_{2},\ldots, \tilde{\mathbf{x}}_{L}\}$ or compactly $\mathbf{x}_{0:L}[\tilde{\mathbf{x}}_{0}]$ is obtained by propagating the equations \eqref{telescopicEqualities1}-\eqref{telescopicEqualitiesL} from $\tilde{\mathbf{x}}_{0}$ in the case FE dynamics or by solving these equations starting from  $\tilde{\mathbf{x}}_{0}$ in the case of the TI dynamics. It should be noted that this cost function is similar to the cost functions in  \eqref{OptimizationProblem1} and \eqref{OptimizationProblem1Relaxed}, except for the $\ell_{1}$ norm and substituted initial state. We determine optimal sensor locations as the solution of the following optimization problem:
\begin{align}
& \min_{\boldsymbol{\theta}} J[\boldsymbol{\theta}],  \label{costFunction00} \\
& \text{subject to} \;\; \sum_{l=1}^{N} \theta_{l} \le M_{\text{max}}, \; \Big(\sum_{l=1}^{N} \theta_{l} = M_{\text{max}} \Big), \label{costFunction00constraint201}\\
& \theta_{l}\in \{0,1 \},\; l=1,2,\ldots,N. \label{costFunction00constraint2}  
\end{align}

To solve \eqref{costFunction00}-\eqref{costFunction00constraint2}, we first transform this problem into an MILP problem. By using the parametrization of the matrix $C_{\boldsymbol{\theta}}[\boldsymbol{\theta}]$ introduced in Section~\ref{problemFormulation}, we can transform the cost function \eqref{costFunction00} into the following form

\begin{align}
J[\boldsymbol{\theta}]= \sum_{j=0}^{L} \sum_{i=1}^{N} \big|z^{(i)}_{j}-\theta_{i}\tilde{x}_{j1}^{(i)}  \big|,
\label{costFunctionTransformation}
\end{align}

where $z^{(i)}_{j}$ is the $i$-th entry of $\mathbf{z}_{j}$, and $\tilde{x}_{j1}^{(i)}$ is the first entry of $\tilde{\mathbf{x}}^{(i)}_{j}$, and this vector is the $i$-th entry of $\tilde{\mathbf{x}}_{j}$ (the index $j$ is the discrete-time instant). By introducing the slack variable vector $\mathbf{b}\in \mathbb{R}^{N(L+1)}$, and by taking into account \eqref{costFunctionTransformation}, we can transform the optimization problem \eqref{costFunction00}-\eqref{costFunction00constraint2} as follows

\begin{subequations}\label{problemAbsoluteValues00}
{
%\begin{empheq}[box=\fbox]{align}
\boxalign[0.43\textwidth]{
\begin{align}
&(\textbf{MILP1})\;\min_{ \mathbf{b},\boldsymbol{\theta}}  \sum_{j=0}^{L} \sum_{i=1}^{N}   b_{j}^{(i)},
\label{OptimizationProblemAbsolute00} \\
&\text{subject to}\;\; \notag \\
& z^{(i)}_{j}-\theta_{i}\tilde{x}_{j1}^{(i)} \le b_{j}^{(i)},\;\;  \theta_{i}\tilde{x}_{j1}^{(i)}- z^{(i)}_{j} \le b_{j}^{(i)}, b_{j}^{(i)} \ge 0, \label{constraint1Absolute00} \\
&   \sum_{i=1}^{N} \theta_{i} \le M_{\text{max}} , ( \sum_{i=1}^{N} \theta_{i} = M_{\text{max}}), \;\theta_{i}\in \{0,1 \},\;         \label{constraint2Absolute00}  \\
& i=1,\ldots, N, \; j=0,1,\ldots, L,\notag
\end{align}
}
%\end{empheq}
}
\end{subequations}
where $b_{j}^{(i)}$ is the $i$-th entry of $\mathbf{b}_{j}$, and $\mathbf{b}_{j}$ is the $j$-th entry of $\mathbf{b}=\text{col}(\mathbf{b}_{0},\mathbf{b}_{1},\ldots, \mathbf{b}_{L})$.

The problem \eqref{problemAbsoluteValues00} is a MILP problem that we solve using the MATLAB function $\text{intlinprog}(\cdot)$. The MATLAB function $\text{intlinprog}(\cdot)$ implements a hybrid method partly based on a branch and bound method. Algorithm~\ref{algorithm2} summarizes this approach for approximating the solution of the MINLP problem.

	\begin{algorithm}[h]
	\caption{Sensor Selection by Solving the Relaxed Problem \textbf{P2} and \textbf{MILP1}}\label{algorithm2}
%		\DontPrintSemicolon
				\textbf{inputs:} In the first phase (steps 1-3), the inputs are the output sequence $\{\mathbf{z}_{0}, \mathbf{z}_{1}, \ldots, \mathbf{z}_{L} \}$ and $M_{\text{max}}$. In the second phase (step 4), the input is the output sequence $\{\mathbf{y}_{0}, \mathbf{y}_{1}, \ldots, \mathbf{y}_{L} \}$ collected by the assigned sensor nodes. \\		
		\textbf{outputs:} The optimal sensor selection vector $\hat{\boldsymbol{\theta}}$ and the initial state estimate $\hat{\mathbf{x}}_{0}$. \\
		\textbf{1. initial solution:} Solve the relaxed problem \textbf{P2} in \eqref{problemPIrelaxed}. Let the solution of this problem be denoted by $(\tilde{\boldsymbol{\theta}},\tilde{\mathbf{x}}_{0})$. \\ 
		\textbf{2. compute state sequence:} Using $\tilde{\mathbf{x}}_{0}$  compute the state sequence $\{\tilde{\mathbf{x}}_{1},\tilde{\mathbf{x}}_{2},\ldots, \tilde{\mathbf{x}}_{L}\}$ by solving the telescopic equations \eqref{telescopicEqualities1}-\eqref{telescopicEqualitiesL}.
		\\
		\textbf{3. solve:} Using $\{\tilde{\mathbf{x}}_{0},\tilde{\mathbf{x}}_{1},\ldots, \tilde{\mathbf{x}}_{L}\}$ form and solve \eqref{problemAbsoluteValues00}. Let the solution of this problem be denoted by  $\hat{\boldsymbol{\theta}}$. \\
		\textbf{4. solve:} Implement the sensor locations on the physical system. Using the installed sensors, collect the output sequence $\{\mathbf{y}_{0}, \mathbf{y}_{1}, \ldots, \mathbf{y}_{L} \}$. Form the matrix $\hat{C}$ by eliminating the zero rows of the matrix $C_{\boldsymbol{\theta}}(\hat{\boldsymbol{\theta}})$ in \eqref{parametrizedOutputEquation}. Setting $C := \hat{C}$, solve \eqref{nonlinearOptimizationProblem} to compute $\hat{\mathbf{x}}_{0}$.			
\end{algorithm}

\subsection{Algorithm 3 - Solving Relaxed and Binary MILP Problem}
Algorithm~\ref{algorithm2} is based on computing the state sequence once the relaxed problem is solved. Due to the specific structure of the cost function \eqref{costFunction}, we can create another cost function in which the initial state is eliminated. This cost function penalizes the difference between the solution $\tilde{\boldsymbol{\theta}}$ of the relaxed problem \textbf{P2} in \eqref{problemPIrelaxed} and the optimization variable $\boldsymbol{\theta}$. The third algorithm computes optimal sensor locations by minimizing this cost function.
This approach is inspired by an approach for solving mixed-integer optimal control problems in~\cite{sager2009reformulations,sager2011combinatorial,burger2019design}. However, our problem formulation and introduced cost function differ from the ones considered in~\cite{sager2009reformulations,sager2011combinatorial,burger2019design}.

 Consider the expression $G$ in  \eqref{costFunction222}. The difference between the value of $G$ for the relaxed solution $\tilde{\boldsymbol{\theta}}$ and for the binary solution $\boldsymbol{\theta}$ that we want to determine is given by 
\begin{align}
G[\tilde{\boldsymbol{\theta}}]-G[\boldsymbol{\theta}] =  \Big(I_{L+1}\otimes \big(C_{\boldsymbol{\theta}} [\boldsymbol{\theta} ] - C_{\tilde{\boldsymbol{\theta}}} [\tilde{\boldsymbol{\theta}} ]   ) \Big) \mathbf{x}_{0:L}[\tilde{\mathbf{x}}_{0}].
\label{binarySolution1}
\end{align}
Our goal is to find $\boldsymbol{\theta}$ such that an upper bound on the $\ell_{1}$ norm of this difference is minimized while ensuring that the constraints on the total number of sensor nodes are satisfied. From \eqref{binarySolution1} we have

\begin{align}
\left\| G[\tilde{\boldsymbol{\theta}}]-G[\boldsymbol{\theta}]\right\|_{1}\le \left\| I_{L+1}\otimes \big(C_{\boldsymbol{\theta}} [\boldsymbol{\theta} ] - C_{\tilde{\boldsymbol{\theta}}} [\tilde{\boldsymbol{\theta}}]   )   \right\|_{1}\left\|  \mathbf{x}_{0:L}[\tilde{\mathbf{x}}_{0}] \right\|_{1}.
\label{l1difference}
\end{align}

This upper bound can be minimized by minimizing the first term on the right-hand-side of \eqref{l1difference}. This term is minimized if $\left\| C_{\boldsymbol{\theta}} [\boldsymbol{\theta} ] - C_{\tilde{\boldsymbol{\theta}}} [\tilde{\boldsymbol{\theta}} ]  \right\|_{1}$ is minimized. Recalling the definition of the $\ell_{1}$ norm, we conclude that in order to minimize $\left\| C_{\boldsymbol{\theta}} [\boldsymbol{\theta} ] - C_{\tilde{\boldsymbol{\theta}}} [\tilde{\boldsymbol{\theta}} ]  \right\|_{1}$, we need to minimize the maximum of all the column sums of matrix difference $C_{\boldsymbol{\theta}} [\boldsymbol{\theta} ] - C_{\tilde{\boldsymbol{\theta}}} [\tilde{\boldsymbol{\theta}} ] $. Since the matrix $C_{\boldsymbol{\theta}} [\boldsymbol{\theta} ]$ has a (block) diagonal structure and due to the parametrization introduced in Section~\ref{problemFormulation}, the problem boils down to the problem of minimizing the maximal difference between the entries of  $\boldsymbol{\theta}$ and $\tilde{\boldsymbol{\theta}}$.  Consequently, to compute optimal sensor locations, we formulate the following optimization problem:
\begin{align}
& \min_{\boldsymbol{\theta}} \max_{i} | \theta_{i}- \tilde{\theta}_{i}  |, \label{costFunction11} \\
& \text{subject to} \;\; \sum_{l=1}^{N} \theta_{l} \le M_{\text{max}}, \Big(\sum_{l=1}^{N} \theta_{l}= M_{\text{max}} \Big), \label{costFunction11constraint20}\\ 
&\theta_{l}\in \{0,1 \},\; l=1,2,\ldots,N. \label{costFunction11constraint21}  
\end{align}

By introducing the slack variable $q\in \mathbb{R}^{+}_{0}$, the last optimization problem can be reformulated as follows
\begin{subequations}\label{problemAbsoluteValues22}
{
%\begin{empheq}[box=\fbox]{align}
\boxalign[0.43\textwidth]{
\begin{align}
&(\textbf{MILP2})\;\;\;\; \min_{ q,\boldsymbol{\theta}}  q     ,
\label{OptimizationProblemAbsolute122} \\
&\text{subject to}\;\; \theta_{i}- \tilde{\theta}_{i} \le q, \;\; \tilde{\theta}_{i}- \theta_{i} \le q,\; i=1,2,\ldots, N,\label{constraint1Absolute22} \\
&   \sum_{l=1}^{N} \theta_{l} \le M_{\text{max}},  \Big(\sum_{l=1}^{N} \theta_{l}= M_{\text{max}} \Big), \label{constraint2Absolute220} \\
& \theta_{l}\in \{0,1 \},\; l=1,2,\ldots,N.              \label{constraint2Absolute221}  
\end{align}
}
%\end{empheq}
}
\end{subequations}

Similarly to the \textbf{MILP1} problem in \eqref{problemAbsoluteValues00}, the \textbf{MILP2} problem in ~\eqref{problemAbsoluteValues22} is solved by using the hybrid method implemented in the MATLAB function $\text{intlinprog}(\cdot)$. Algorithm~\ref{algorithm3} summarizes the third approach for selecting the sensor nodes.

	\begin{algorithm}[h]
	\caption{Sensor Selection by Solving Relaxed Problem \textbf{P2} and \textbf{MILP2}}\label{algorithm3}
%		\DontPrintSemicolon
		\textbf{inputs:} In the first phase (steps 1-2), the inputs are the output sequence $\{\mathbf{z}_{0}, \mathbf{z}_{1}, \ldots, \mathbf{z}_{L} \}$ and the maximal number of sensor nodes $M_{\text{max}}$. In the second phase (step 3), the inputs is the output sequence $\{\mathbf{y}_{0}, \mathbf{y}_{1}, \ldots, \mathbf{y}_{L} \}$ collected by the assigned sensor nodes. \\	
		\textbf{outputs:} The optimal sensor selection vector $\hat{\boldsymbol{\theta}}$ and the initial state estimate $\hat{\mathbf{x}}_{0}$. \\
		\textbf{1. initial solution:} Solve the relaxed problem P2 in \eqref{problemPIrelaxed}. Let the solution of this problem be denoted by $(\tilde{\boldsymbol{\theta}},\tilde{\mathbf{x}}_{0}^{(0)})$. \\
		\textbf{2. solve:} Using $\tilde{\boldsymbol{\theta}}$  solve \eqref{problemAbsoluteValues22}. Let the solution of this problem be denoted by  $\hat{\boldsymbol{\theta}}$ \\
		\textbf{3. solve:} Implement the sensor locations on the physical system. Using the installed sensors, collect the output sequence $\{\mathbf{y}_{0}, \mathbf{y}_{1}, \ldots, \mathbf{y}_{L} \}$. Form the matrix $\hat{C}$ by eliminating the zero rows of the matrix $C_{\boldsymbol{\theta}}(\hat{\boldsymbol{\theta}})$. Setting $C := \hat{C}$, solve \eqref{nonlinearOptimizationProblem} to compute $\hat{\mathbf{x}}_{0}$.		
\end{algorithm}

In the general case, the formulated MINO problem \textbf{P1} in \eqref{problemPInteger} is NP-hard, non-convex, and constrained by nonlinear dynamics. On the other hand, the introduced methods are approximating the solution. Due to the many layers of complexity, theoretical investigation of the degree of the suboptimality of the developed methods is challenging. To provide crucial insights into the optimality of the developed approaches, in Section~\ref{numericalResults} we investigate the performance of our methods by performing extensive numerical experiments. The good numerical performance of our methods is demonstrated by testing the methods on Duffing oscillator networks that are prototypical models of a number of dynamical systems, as well as on associative memory networks representing simplified memory models and on chemical reaction networks. We compare our methods with exhaustive search and random selection of sensor nodes.

\section{Derivatives and Generalizations}
\label{gradientsAndGeneralizations}

To significantly speed up the computations of the solutions of the optimization problems, in this section we derive expressions for gradients and derivatives of the cost function used in the problems~\eqref{problemPInteger} and \eqref{problemPIrelaxed}. In addition, in Section~\ref{generalizationSubsections} we briefly present generalizations of the optimization problems \eqref{problemPInteger} and \eqref{problemPIrelaxed} that incorporate a number of output and state sequences.  In this way, we can increase the robustness and decrease the sensitivity of the sensor selection procedure with respect to uncertainties and lack of knowledge of the initial state of the system.

\subsection{Derivatives}
First, we start with the derivatives of the FE and TI dynamics in \eqref{FEdiscretizedDynamics} and \eqref{TEdiscretizedDynamics}. Similar to the recursive computation (simulation) of state sequences, derivatives are also computed in a recursive manner. From \eqref{FEdiscretizedDynamics} it follows that for the FE dynamics, we have 
\begin{align}
\frac{\partial \mathbf{x}_{k}}{\partial \mathbf{x}_{0}}=\frac{\partial \mathbf{x}_{k-1}}{\partial \mathbf{x}_{0}}+h\frac{\partial \mathbf{x}_{k-1}}{\partial \mathbf{x}_{0}}\frac{\partial \mathbf{f}[ \mathbf{x}_{k-1}]}{\partial \mathbf{x}_{k-1}}.
\label{FEdynamics22}
\end{align}
On the other hand, from \eqref{TEdiscretizedDynamics} it follows that for the TI dynamics, we have
\begin{align}
\frac{\partial \mathbf{x}_{k}}{\partial \mathbf{x}_{0}}=\frac{\partial \mathbf{x}_{k-1}}{\partial \mathbf{x}_{0}}+ \frac{h}{2} \Big( \frac{\partial \mathbf{x}_{k}}{\partial \mathbf{x}_{0}}\frac{\partial \mathbf{f}[ \mathbf{x}_{k}]}{\partial \mathbf{x}_{k}} + \frac{\partial \mathbf{x}_{k-1}}{\partial \mathbf{x}_{0}}\frac{\partial \mathbf{f}[ \mathbf{x}_{k-1}]}{\partial \mathbf{x}_{k-1}}
 \Big). \label{TIdynamicsPropagation22}
\end{align}
Under a mild assumption of invertibility of $\big(I_{Nn}-(h/2)\partial \mathbf{f}[ \mathbf{x}_{k}]/\partial \mathbf{x}_{k}\big)$, from the last expression, we obtain
\begin{align}
\frac{\partial \mathbf{x}_{k}}{\partial \mathbf{x}_{0}}=\frac{\partial \mathbf{x}_{k-1}}{\partial \mathbf{x}_{0}}\Big(I_{Nn}+\frac{h}{2}\frac{\partial \mathbf{f}[ \mathbf{x}_{k-1}]}{\partial \mathbf{x}_{k-1}}  \Big)\Big(I_{Nn}-\frac{h}{2}\frac{\partial \mathbf{f}[ \mathbf{x}_{k}]}{\partial \mathbf{x}_{k}}  \Big)^{-1}. 
\label{TIdynamicsPropagation221}
\end{align}
By initializing \eqref{FEdynamics22} and \eqref{TIdynamicsPropagation221} with $\partial \mathbf{x}_{0} / \partial \mathbf{x}_{0}= I_{Nn} $, we can recursively compute the derivatives of state sequences. These derivatives will be used to compute the gradients of the cost function
\begin{align}
\mathcal{W}[\boldsymbol{\theta},\mathbf{x}_{0}]=\left\| \mathbf{z}_{0:L}-K[\boldsymbol{\theta}] \mathbf{x}_{0:L}[\mathbf{x}_{0}]\right\|_{2}^{2},\; K[\boldsymbol{\theta}]=  \big(I_{L+1}\otimes C_{\boldsymbol{\theta}}[\boldsymbol{\theta}] \big), \label{costFunctionRedefinition}
\end{align}
that appear in the MINO problem \textbf{P1} defined in \eqref{problemPInteger} and in the relaxed problem \textbf{P2} defined in  \eqref{problemPIrelaxed}. 
The gradient of the cost function is defined by 
\begin{align}
\nabla \mathcal{W} =\begin{bmatrix} \nabla_{\mathbf{x}_{0}} \mathcal{W} \\ \nabla_{\boldsymbol{\theta}} \mathcal{W} \end{bmatrix}.
\label{gradients}
\end{align}
For brevity we only give the final gradient expressions. We obtained
\begin{align}
 &\nabla_{\mathbf{x}_{0}} \mathcal{W}=-2\frac{\partial \mathbf{x}_{0:L}[\mathbf{x}_{0}] }{\partial  \mathbf{x}_{0}}K^{T}[\boldsymbol{\theta}]\big(\mathbf{z}_{0:L}-K[\boldsymbol{\theta}] \mathbf{x}_{0:L}[\mathbf{x}_{0}] \big), \label{tmp1gradient} \\
& \frac{\partial \mathbf{x}_{0:L}[\mathbf{x}_{0}] }{\partial  \mathbf{x}_{0}}=\begin{bmatrix}  I_{Nn} & \frac{\partial \mathbf{x}_{1}}{\partial \mathbf{x}_{0}} & \ldots &  \frac{\partial \mathbf{x}_{k}}{\partial \mathbf{x}_{0}} \end{bmatrix}, \label{tmp2gradient} \\
& \nabla_{\boldsymbol{\theta}} \mathcal{W}=-2\begin{bmatrix} D_{0} &  D_{1} & \ldots & D_{L} \end{bmatrix}\Big( \mathbf{z}_{0:L}-K[\boldsymbol{\theta}] \mathbf{x}_{0:L}[\mathbf{x}_{0}]\Big),
\label{tmp3gradient}
\end{align}
where $D_{i}=\text{diag}(C_{1} \mathbf{x}_{i})$ and the matrix $C_{1}$ depends on the parametrization of $C_{\boldsymbol{\theta}}[\boldsymbol{\theta}]$.  In the case when $C_{\boldsymbol{\theta}}[\boldsymbol{\theta}]=\text{diag}(\boldsymbol{\theta})$, we have $C_{1}=\text{diag}(\mathbf{1})$, where $C_{1}\in \mathbb{R}^{N\times N}$. In the case when $C_{\boldsymbol{\theta}}[\boldsymbol{\theta}]$ is a block diagonal matrix with the $i$-th block $[\theta_{1} \; 0 \; \ldots 0]\in \mathbb{R}^{1\times n}$, the matrix $C_{1}\in \mathbb{R}^{N\times Nn}$ is a block diagonal matrix with the $i$-th block $[1 \; 0 \; \ldots 0]\in \mathbb{R}^{1\times n}$. To compute \eqref{tmp1gradient} and \eqref{tmp2gradient}, we use the recursive expressions \eqref{FEdynamics22} and \eqref{TIdynamicsPropagation221}.

\subsection{Generalizations}
\label{generalizationSubsections}

The sensor selection approach presented in Section~\ref{sensorSelectionAlgorithms} is derived on the basis of the output sequence $\{\mathbf{z}_{k}\}$. In the first sensor selection scenario, explained in Section~\ref{problemFormulation}, this sequence is computed by simulating the dynamics for the user-selected initial condition. Due to the nonlinearity and non-convexity of the problem, the computed selections of sensor nodes might depend to some extent on the user-selected initial condition.  It might happen that the selected sensor locations do not produce satisfactory results when the initial state of the real system significantly differs from the one used to generate the data sequence $\{\mathbf{z}_{k}\}$. In addition, the solutions of the introduced optimization problems might depend on the initial guesses of the states and sensor locations. If we know the operating range of the initial states of the real system, then we can use this knowledge to properly choose the user-defined initial state for generating the sequence $\{\mathbf{z}_{k}\}$ and for initializing the optimization problems. In this way, we can reduce the sensitivity of the sensor selection procedure with respect to user-selected initial states. However, in some cases, due to the lack of knowledge, this is impossible. Our approach for dealing with this problem is to generalize the cost functions in \eqref{problemPInteger} and \eqref{problemPIrelaxed}, by including many output and state sequences $\{ \mathbf{z}_{k} \}$ generated for different selections of initial conditions. We can replace the cost functions defined in Section~\ref{sensorSelectionAlgorithms} by the following cost function
\begin{align}
& W[\boldsymbol{\theta},\mathbf{x}_{0}]=\left\| Z-K[\boldsymbol{\theta}] X\right\|_{F}^{2},
\label{generalizedCostFunctions} \\
& Z=\begin{bmatrix}\mathbf{z}_{0:L,1} & \mathbf{z}_{0:L,2} & \ldots & \mathbf{z}_{0:L,P} 
 \end{bmatrix},\notag \\
& X=\begin{bmatrix} \mathbf{x}_{0:L,1}[\mathbf{x}_{0,1}] & \mathbf{x}_{0:L,2}[\mathbf{x}_{0,2}] & \ldots & \mathbf{x}_{0:L,P}[\mathbf{x}_{0,P}] \end{bmatrix},
\end{align}
where $\mathbf{z}_{0:L,i}$ and $ \mathbf{x}_{0:L,i}$, $i=1,2,\ldots, P$, are the output and state sequences, respectively, $K[\boldsymbol{\theta}]$ is defined in \eqref{costFunctionRedefinition}, and $\left\| \cdot \right\|_{F}^{2}$ is the Frobenius norm. The sequences $\{\mathbf{z}_{0:L,i} \}$ are obtained by simulating the system dynamics for different initial states. The number $P$ is the total number of user-selected initial states. For example, the initial states can be  selected as corners of a convex hull of the expected state-space region from where the state trajectory starts. In some sense, this generalization produces an averaged sensor selection $\hat{\boldsymbol{\theta}}$ that is optimal for a number of initial conditions. The disadvantage of this approach is that the computational complexity of solving the sensor selection problems is increased.

\section{Numerical Results}
\label{numericalResults}
In this section, we present numerical results of applying the developed algorithms to associative memory, Duffing oscillator, and chemical reaction networks. Due to the paper brevity, we mainly focus on networks with stable equilibrium points, and in future work, we will perform tests on networks with limit-cycle dynamics. The simulations are performed on a computer with 16GB RAM and Intel Core i7-8700 processor. The developed codes are available online~\cite{haberSensorCode2020}.

\subsection{Sensor Selection for Associative Memory Networks}
Associative memory networks can be seen as simplified memory models~\cite{haber2020submitted,cornelius2013realistic,nishikawa2004capacity}. The idea is to select network parameters such that starting from an initial state, representing a perturbed letter or a binary pattern, the network state converges to the equilibrium state, representing the memorized letter or the memorized binary pattern. For brevity, we give a final state-space form of the network, for more background information, see~\cite{haber2020submitted,cornelius2013realistic,nishikawa2004capacity} and references therein. The dynamics of the $i$-th node has the following form
\begin{align}
\dot{x}_{i}=\sum_{j=1}^{N} \beta_{ij} \text{sin}\big(x_{j}- x_{i}\big)+\frac{\gamma }{N} \sum_{j=1}^{N}\text{sin}2\big(x_{j}- x_{i}\big),
\label{memoryNetworks}
\end{align}
where $x_{i}\in \mathbb{R}$, $\gamma =0.8$, $\beta_{ij}\in \mathbb{R}$ is given by $\beta_{ij}=(1/N) \sum_{\omega =1}^{p} \zeta_{i}^{\omega}\zeta_{j}^{\omega}$ (Hebb's learning rule), where  $\zeta_{i}^{\omega}=\pm 1$, $\omega=1,2,\ldots, p$, and $p$ denotes the number of binary patterns to be memorized. A binary pattern to be memorized is represented by the vector  $\boldsymbol{\zeta}^{\omega}=\mathrm{col}\big( \zeta_{1}^{\omega}, \zeta_{2}^{\omega},\ldots, \zeta_{N}^{\omega} \big)$. Let $\boldsymbol{\zeta}^{1}, \boldsymbol{\zeta}^{2},\ldots, \boldsymbol{\zeta}^{p}$ represent $p$ desired binary patterns that need to be memorized by the network. Then for such a selection of binary patterns, we can compute $\beta_{ij}$ using the previously explained formula, and we can construct the network dynamics~\eqref{memoryNetworks}. 
\begin{figure}[t]
	\centering 
	\includegraphics[scale=0.37,trim=0mm 0mm 0mm 0mm ,clip=true]{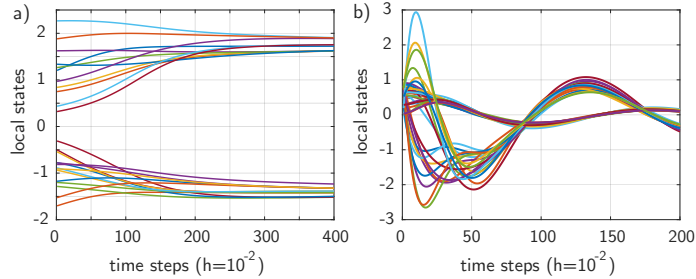}
	\caption{The state trajectories of a) the associative memory networks and b) Duffing oscillator networks.}
	\label{fig:Graph01}
\end{figure}
We use $N=25$ and we choose the FE method to discretize the dynamics since as demonstrated in~\cite{haber2020submitted} this method produces a small discretization error and the network dynamics is not stiff. We define binary patterns on a $5\times 5$ grid, corresponding to the letters  ``H", ``T", and ``L" ($p=3$). The local state trajectories simulated from a perturbed state are shown in Fig.~\ref{fig:Graph01}(a). In order to be in the transient regime during sensor selection, we use $h=10^{-3}$ and $L=21$.  
The network initial state is the ``T" letter that is perturbed by the normal Gaussian noise. Our goal is to estimate this initial state. We test the performance of the method for $40\%,56\%,72\%,$ and $88\%$ of observed nodes (corresponding to $M_{\text{max}}=10,14,18,22$). For Algorithm~\ref{algorithm1}, we use the less than equal constraints in \eqref{problemPInteger}. This is necessary since we have noticed that in the case of equality constraints the NOMAD solver occasionally produces infeasible solutions. On the other hand, in the case of Algorithms~\ref{algorithm2} and \ref{algorithm3}, we use the equality constraints. For the relaxed problem \textbf{P2} in \eqref{problemPIrelaxed}, the initial guess of the unknown state is a vector with entries drawn from the Gaussian normal distribution, and the initial guess of the sensor locations is a vector with entries drawn from the uniform distribution on $[0,1]$. In the relaxed problem, we use the lower and upper bounds equal to $-5$ and $5$, respectively, and $0$ and $1$ for the relaxed binary constraints.

\begin{figure}[t]
	\centering 
	\includegraphics[scale=0.53,trim=0mm 0mm 0mm 0mm ,clip=true]{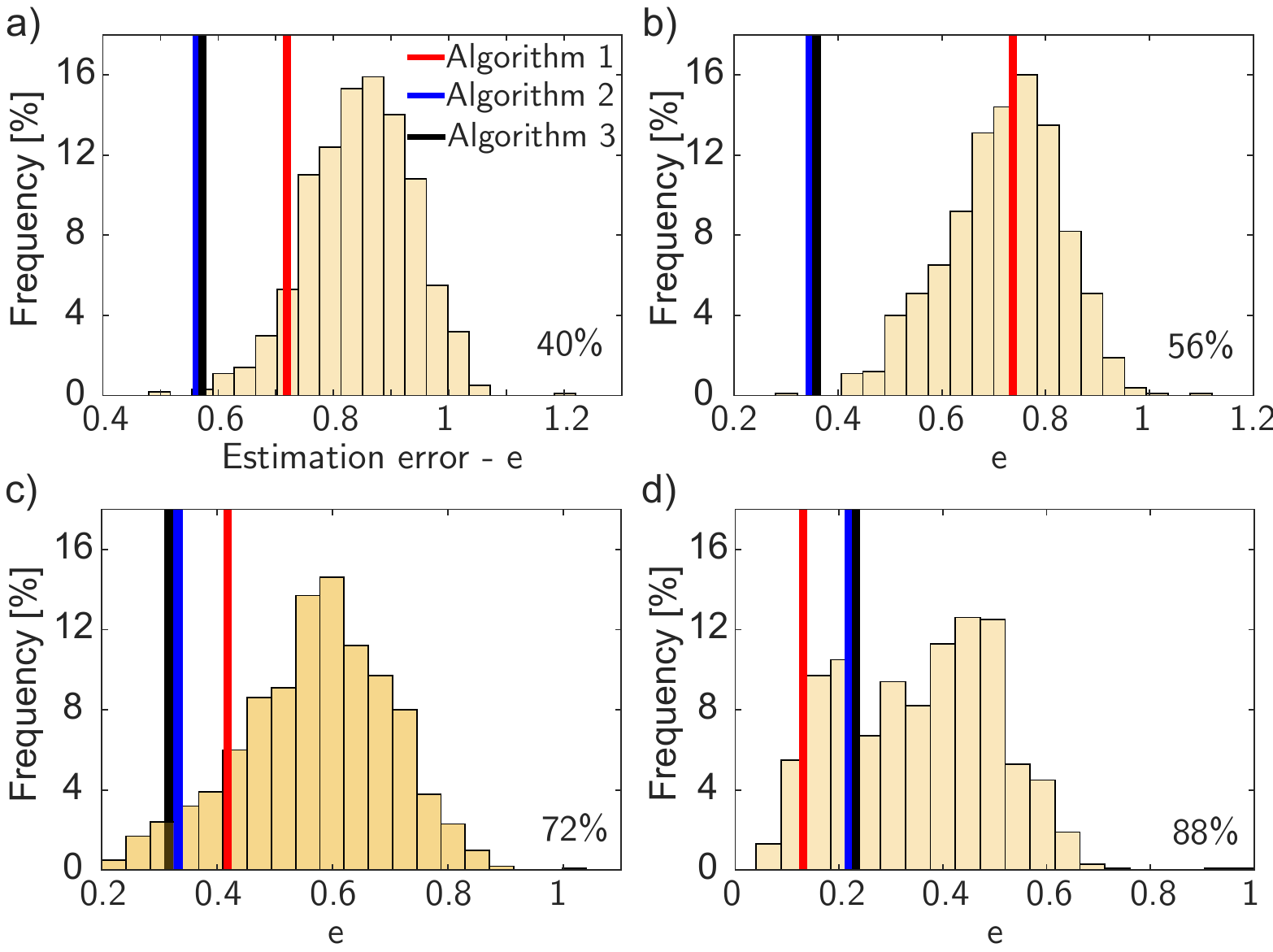}
	\caption{The estimation errors for associative memory networks. The vertical colored lines correspond to the final estimation errors produced by the developed algorithms and histograms correspond to empirical error distributions obtained by random sensor selection for the fixed fraction of observation nodes. (a) $40 \%$, (b) $56 \%$, $72 \%$, and $88 \%$ of observed nodes. }
	\label{fig:Graph1}
\end{figure}

To test the performance of the developed algorithms, we compare optimal sensor selections with a sequence of random selections of control nodes. For every random selection, we form the corresponding matrix $C$, and we estimate the initial state by solving \eqref{nonlinearOptimizationProblem}. The estimation error is quantified by $e=\left\|\mathbf{x}_{0}^{\text{true}}-\hat{\mathbf{x}}_{0} \right\|_{2} /\left\| \mathbf{x}_{0}^{\text{true}} \right\|_{2} $, where $\mathbf{x}_{0}^{\text{true}}$ is the ``true'' value of the initial state. Figure~\ref{fig:Graph1} compares the performance of the developed methods with random sensor selections. We have generated 1000 random selections of sensor nodes. Histograms correspond to empirical distributions of the estimation error produced by random sensor selections. The colored vertical lines denote estimation errors produced by optimal sensor selections computed by the developed algorithms. We can notice that the performance of algorithms varies with a fraction of observed nodes. However, for a fixed fraction of sensor nodes, there is at least one algorithm that produces an error that is smaller than most of the errors produced by the random sensor selection. This is a numerical verification of a relatively good performance of the developed approaches.

\subsection{Sensor Selection for Duffing Oscillator Networks}

Duffing oscillators are dynamical systems with a nonlinear spring stiffness $F_{s}=\eta x - \chi x^{3}$, where $x$ is the spring displacement, $\eta,\chi \in \mathbb{R}$ are the spring constants, and $F_{s}$ is the spring force (we assume a softening spring). These oscillators are prototypical models of a large number of dynamical systems, such as electrical circuits, nanomechanical resonators, structural beams, cables, etc. An excellent introduction to the theory and applications of Duffing oscillators is given in~\cite{kovacic2011duffing}. We consider a nonlinear network consisting of damped Duffing oscillators with nonlinear connections:
\begin{align}
& \dot{x}_{i1} = x_{i2},  \label{localModelDuffingState1}  \\
&\dot{x}_{i2}= -\eta_{ii} x_{i1} +\chi_{ii}x_{i1}^{3}-\rho_{ii}x_{i2}-\sum_{j\in \mathcal{N}(i)} \eta_{ij}\big(x_{i1}-x_{j1}\big) \notag \\ &
+ \sum_{j\in \mathcal{N}(i)} \chi_{ij}\big(x_{i1}-x_{j1}\big)^{3}- \sum_{j\in \mathcal{N}(i)}\rho_{ij}\big(x_{i2}-x_{j2}\big),  \label{localModelDuffingState2} 
\end{align}
where $x_{i1}$ is the displacement and $x_{i2}$ is the velocity of the $i$-th oscillator, $\eta_{ij}$ and $\chi_{ij}$ are previously introduced spring constants, $\rho_{ij}$ is a damping parameter, $\mathcal{N}(i)$ is a set of oscillators that are connected to the $i$-th oscillator. We assume that connections between oscillators are defined by a Geometric Random Graph (GRG). We use the method and codes available in~\cite{taylor2009contest} to generate GRGs. GRGs are constructed by randomly placing nodes on a unit square, and connecting nodes according to a user-defined connection radius. In our simulations, we use the radius of $\sqrt{1.44/N}$. The parameters $\eta_{ij}$ are selected from the uniform distribution defined on the interval $[10,20]$. On the other hand, parameters $\chi_{ij},\rho_{ij}$ are selected from the uniform distribution defined on $[1,2]$. The state of the $i$-th node is $\mathbf{x}^{(i)}=[x_{1i} \;\; x_{2i}]^{T}$. Typical local state trajectories simulated from a random initial state and for $N=20$ are shown in Fig.~\ref{fig:Graph01}(b).

\begin{figure}[t]
	\centering 
	\includegraphics[scale=0.28,trim=0mm 0mm 0mm 0mm ,clip=true]{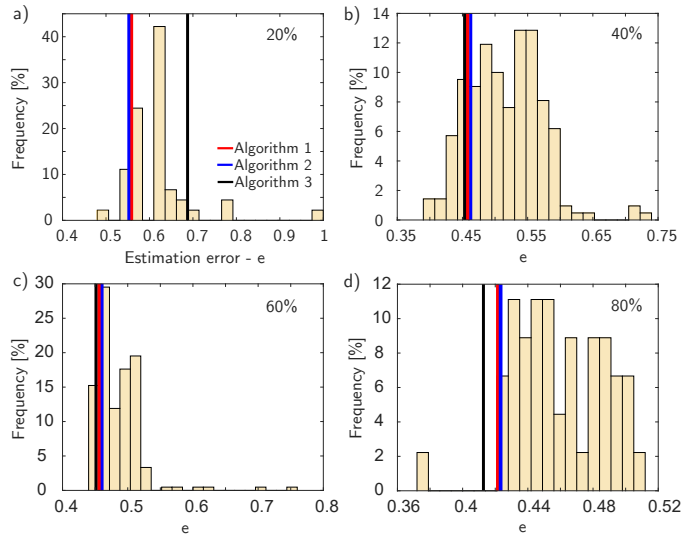}
	\caption{The estimation errors for the Duffing oscillator networks. The interpretation of vertical lines and histograms is the same as in Fig.~\ref{fig:Graph1} and is explained in the caption of Fig.~\ref{fig:Graph1}. (a) $20 \%$, (b) $40 \%$, $60 \%$, and $80 \%$ of observed nodes.}
	\label{fig:Graph2}
\end{figure}

To test the sensor selection procedure, we construct a Duffing oscillator network with $N=10$ nodes. Since $N$ is not large, we can compare our methods with an exhaustive search. In this way, we can quantify how far are computed solutions from the most optimal ones. In our simulations, we choose $h=10^{-3}$ and $L=101$, such that the collected data belongs to the transient regime. We use the TI method since the FE methods for the Duffing network produces unstable results. The entries of a ``true'' initial state (to be estimated) are generated from the uniform distribution on the interval $[0,1]$. An initial guess for this solution is generated using the same principle, however, we make sure that the entries of the guess and ``true'' initial state are different. The initial guess for the sensor nodes for the relaxed problem \textbf{P2} in \eqref{problemPIrelaxed} is generated as a random vector whose entries are selected from the uniform distribution on $[0,1]$. The principle for selecting other optimization parameters is similar to the principle used to select the optimization parameters for the case of memory networks, for more details, see the codes that are posted online~\cite{haberSensorCode2020}. We test the performance of the methods for $20\%,40\%,60\%$, and $80\%$ of observed nodes (corresponding to $M_{\text{max}}=2,4,6,8$). At the same time, we perform the exhaustive search by generating all the possible combinations for fixed fractions of observation nodes, and by estimating an initial state for every combination by solving \eqref{nonlinearOptimizationProblem}. The results are shown in Fig.~\ref{fig:Graph2}. We can observe that for every case of the fixed fraction of observed nodes, at least one algorithm produces the estimation error that is relatively close to the most optimal one. 

\subsection{Sensor Selection for Chemical Reaction Networks}
Due to the length constraints of this manuscript, we provide a brief description of combustion reaction networks, for more details see~\cite{turns1996introduction,smirnov2014modeling,haber2017state} and references therein. We consider a network consisting of $\mathcal{N}$ chemical reactions
\begin{align}
	\sum_{i=1}^{N}\pi_{ji} \mathcal{S}_{i} \rightleftarrows    \sum_{i=1}^{N} \omega_{ji}\mathcal{S}_{i}, \; j=1,2,\ldots, \mathcal{N},
	\label{chemicalReactionNetwork}
\end{align}
where $\mathcal{S}_{i}$, $i=1,2,\ldots, N$ are chemical species, $\pi_{ji}$ and $\omega_{ji}$ are stoichiometric coefficients. In the state-space description of the chemical reaction networks, the states are chemical species concentrations. The state-space model is
\begin{align}
	\dot{\mathbf{x}}(t)=\Phi \boldsymbol{\lambda}[\mathbf{x}],
	\label{continiousTimeFinal}
\end{align}
where $\boldsymbol{\lambda}[\mathbf {x}]=[\lambda_{1}[\mathbf{x}],\lambda_{2}[\mathbf{x}],\ldots, \lambda_{\mathcal{N}}[\mathbf{x}]]^{T}$, and $\Phi=[ \omega_{ji}-\pi_{ji} ]\in \mathbb{R}^{N\times \mathcal{N}}$,  
$\mathbf{x}=[x_{1},x_{2},\ldots, x_{N}]$, and $\lambda_{j}$, $j=1,2,\ldots, \mathcal{N}$ are functions that are polynomials of $\mathbf{x}$. The local states $x_{i}$, $i=1,2,\ldots, N$ are the concentrations of chemical species. The polynomial functions are defined by 
\begin{align}
	\lambda_{j}[\mathbf{x}]= v_{j}^{f}\prod_{i=1}^{N}x_{i}^{\pi_{ji}}-v_{j}^{b}\prod_{i=1}^{N}x_{i}^{\omega_{ji}},\;j=1,2,\ldots, \mathcal{N},
	\label{polynomialQ}
\end{align}
where $v_{j}^{f},v_{j}^{b}\in \mathbb{R}_{+}$ are the forward and backward rate constants computed on the basis of the Arrhenius law. In our simulations, all the network parameters are computed using the Cantera software, for more details, see~\cite{haber2017state} and references therein.

\begin{figure}[t]
	\centering 
	\includegraphics[scale=0.28,trim=0mm 0mm 0mm 0mm ,clip=true]{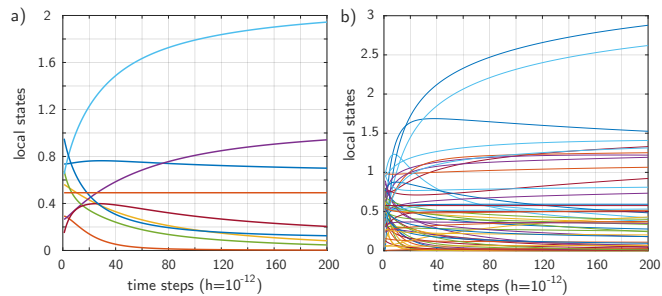}
	\caption{The local state trajectories of the a) $H_{2}/O_{2}$ and b) natural gas combustion networks.}
	\label{fig:Graph02}
\end{figure}

We consider two combustion chemical reaction networks. The first network is the $H_{2}/O_{2}$ network having $9$ species and $27$ reactions (N=9). The second network is the natural gas combustion network GRI-Mech 3.0 consisting of $53$ species and $325$ reactions (N=53). The dynamics is generated for the temperature of $T=1473.15 [K]$ and the pressure of $1$ atmosphere. The local state trajectories of these networks are shown in Fig.~\ref{fig:Graph02}. To discretize the system dynamics, we used the TI method since the networks are relatively stiff. 

\begin{figure}[t]
	\centering 
	\includegraphics[scale=0.30,trim=0mm 0mm 0mm 0mm ,clip=true]{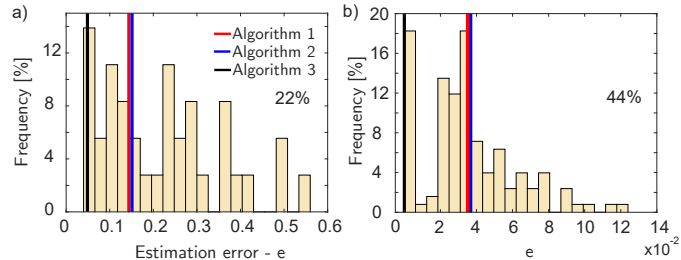}
	\caption{The estimation errors for the $H_{2}/O_{2}$ networks. The interpretation of vertical lines and histograms is the same as in Fig.~\ref{fig:Graph1} and is explained in the caption of Fig.~\ref{fig:Graph1}. (a) $22 \%$ ($M_{\text{max}}=2$) and (b) $22 \%$  ($M_{\text{max}}=4$) of observed nodes.}
	\label{fig:Graph03}
\end{figure}

\begin{figure}[t]
	\centering 
	\includegraphics[scale=0.30,trim=0mm 0mm 0mm 0mm ,clip=true]{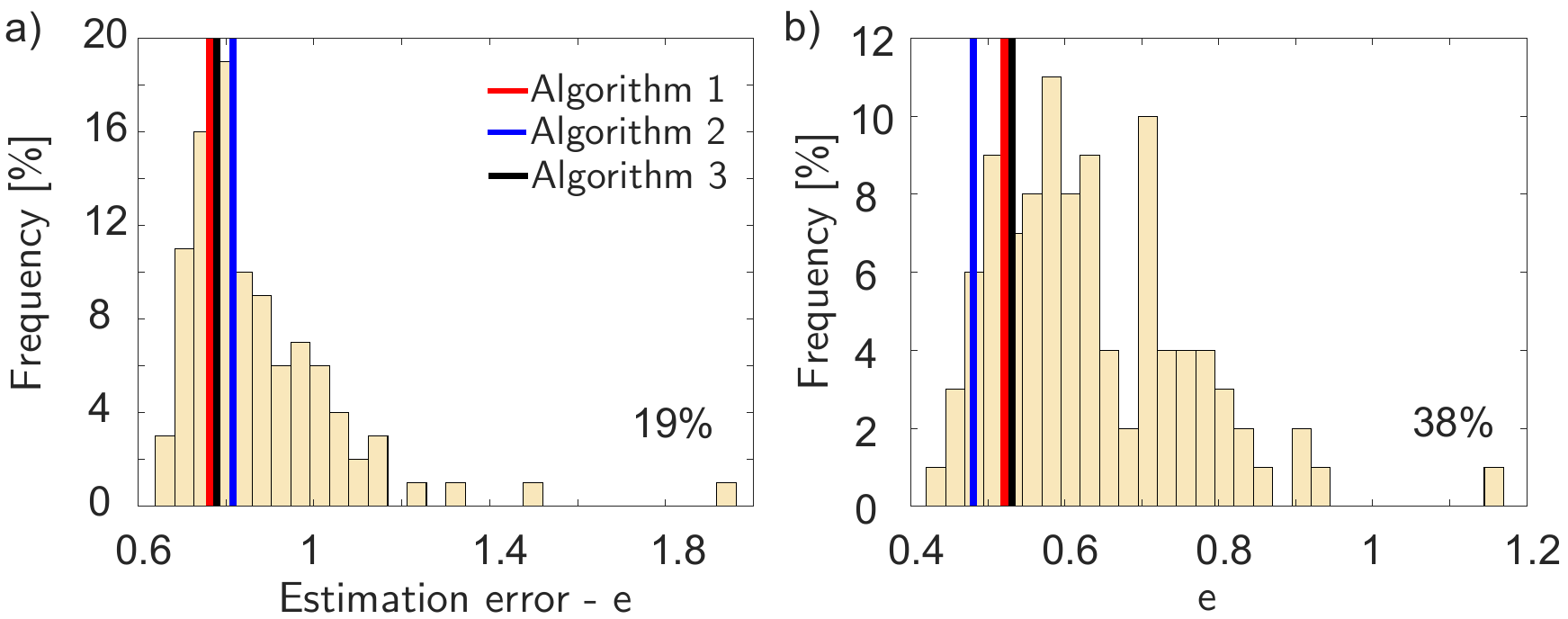}
	\caption{The estimation errors for the GRI 3.0 networks. The interpretation of vertical lines and histograms is the same as in Fig.~\ref{fig:Graph1} and is explained in the caption of Fig.~\ref{fig:Graph1}. (a) $19 \%$ ($M_{\text{max}}=10$) and (b) $38 \%$ ($M_{\text{max}}=20$) of observed nodes.}
	\label{fig:Graph04}
\end{figure}

Figure~\ref{fig:Graph03} and Fig.~\ref{fig:Graph04} show the performance of the developed methods for the $H_{2}/O_{2}$ and natural gas combustion networks, respectively. For both networks, the results are generated for $h=10^{-13}$ and $L=100$. The principle for selecting other optimization parameters is similar to the principle used to select the optimization parameters for the case of memory networks, for more details, see the codes that are posted online~\cite{haberSensorCode2020}. In the case of the $H_{2}/O_{2}$ network we compare the developed methods with an exhaustive search. In the case of the natural gas combustion network, we compare the methods with $100$ random selections of sensor nodes. From Fig.~\ref{fig:Graph03} we can observe that the performance of Algorithm 3 is almost ideal for the case of the $H_{2}/O_{2}$ network. On the other hand, from Fig.~\ref{fig:Graph04} we can observe that the performance of this algorithm is slightly worse for the case of the natural combustion network. The performance of all three algorithms can be improved by increasing the length of the observation horizon. However, this performance gain is at the expense of increasing the computational complexity of the developed approaches.

\subsection{Computational Complexity Tests}

\begin{figure}[t]
	\centering 
	\includegraphics[scale=0.42,trim=0mm 0mm 0mm 0mm ,clip=true]{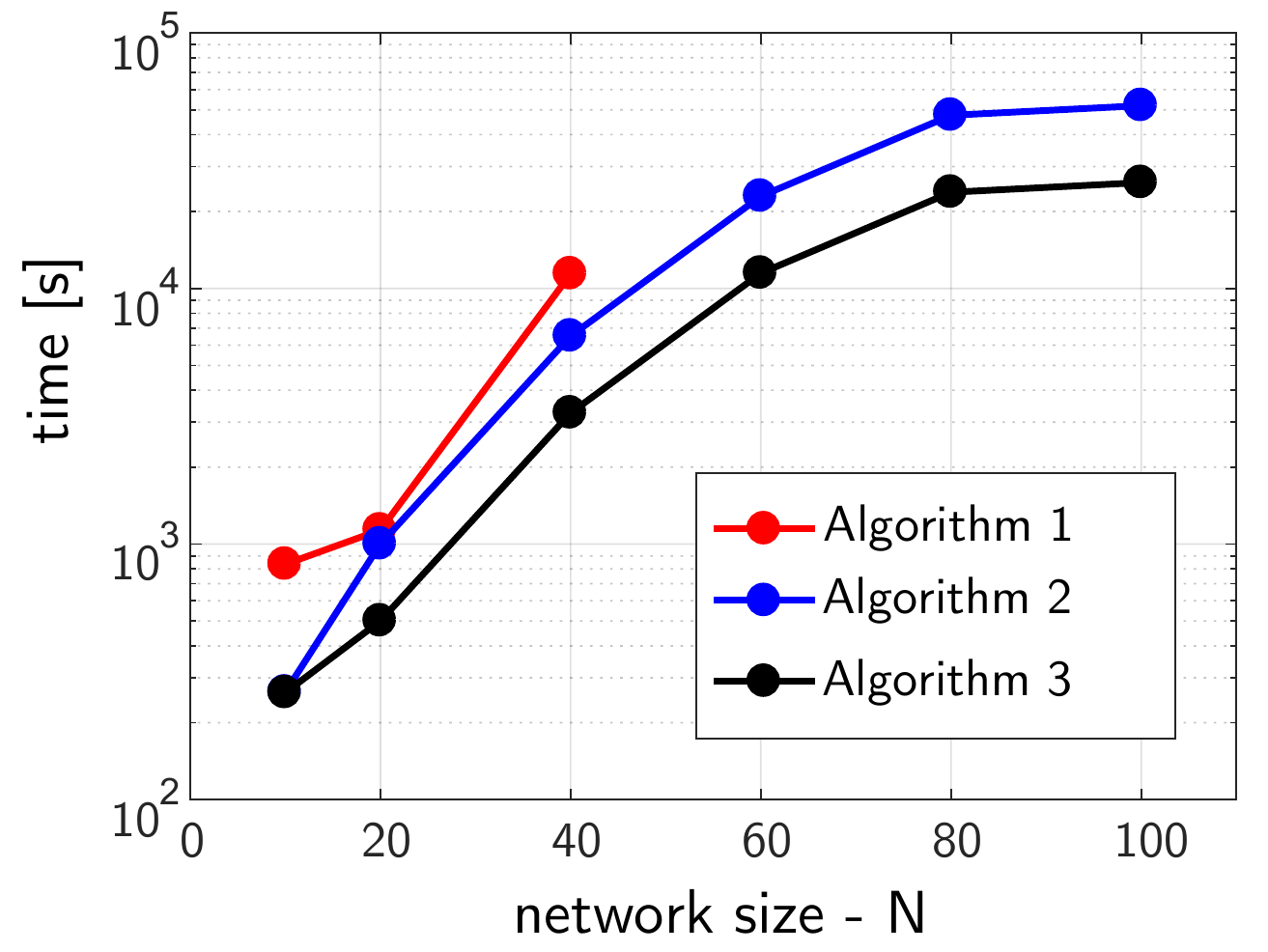}
	\caption{The computational complexity results of the developed algorithms. The results are obtained for the TI dynamics and the Duffing oscillator networks with $L=201$ and $h=10^{-4}$.}
	\label{fig:Graph3}
\end{figure}

Finally, we test the computational complexity of the developed methods on the example of the Duffing oscillator network.  Due to the many layers of complexity and the fact that we use several nonlinear solvers to compute the solutions, it is challenging to state theoretical bounds on the computational complexity. Consequently, we provide numerical insights into computational complexity. The results are shown in Fig.~\ref{fig:Graph3}. Here, we used the TI dynamics, $L=201$, and $h=10^{-4}$. That is, to compute the solution it is necessary to simulate the dynamics by solving the system of nonlinear equations. The computational complexity times are two orders of magnitude smaller for the FE dynamics. From Fig.~\ref{fig:Graph2} we can see that Algorithm~\ref{algorithm3} produces the lowest computational complexity. This is due to the fact that the \textbf{MILP2} problem in \eqref{problemAbsoluteValues22} only contains $N+1$ optimization variables. On the other hand, as expected, Algorithm~\ref{algorithm1} produces the highest computational complexity since it is based on the direct search.  The presented computational complexity results can be improved by employing parallel implementations of the used nonlinear solvers.

\section{Conclusions and Future Work}
\label{conclusions}

In this manuscript, we have developed methods for sensor selection for nonlinear networks. We performed extensive numerical experiments that demonstrate the good performance of the developed methods. We noticed that the performance of the developed approaches largely depends on the parameters such as the length of the observation horizon, discretization constant, and nonlinear solver parameters. Furthermore, we have noticed that the performance of the developed methods depends on the network types. For example, the best results are obtained for the memory and $H_{2}/O_{2}$ networks. In the case of Duffing oscillator and natural gas combustion networks, the performance is slightly worse. The performance can be improved by incorporating higher-order discretization techniques into the developed optimization methods. In future research, we will systematically investigate the influence of network and optimization parameters on the overall performances of the developed methods. On the other hand, due to the nonlinear nature of the problem, optimal sensor selections depend on the output sequence used for optimization. In future research, using the directions explained in Section~\ref{generalizationSubsections}, we will generalize the approach by including a number of output and state sequences in the sensor selection problem and we will investigate different options for reducing the computational complexity of the developed methods. In this way, we will further improve the performance of the developed approaches. In addition, we will test our methods on networks exhibiting limit cycle behavior. Finally, the proposed approaches will serve as the basis for the development of closed-loop observers.

\bibliographystyle{IEEEtran}
\bibliography{sample}

\end{document}